# Robust broad spectral photodetection (UV-NIR) and ultra high responsivity investigated in nanosheets and nanowires of $Bi_2Te_3$ under harsh nano-milling conditions

Alka Sharma[1,2], A. K. Srivastava[1,2], T. D. Senguttuvan[1,2] and Sudhir Husale[1,2]*

[1]Academy of Scientific and Innovative Research (AcSIR), National Physical Laboratory, Council of Scientific and Industrial Research, Dr. K. S Krishnan Road, New Delhi-110012, India.

[2] National Physical Laboratory, Council of Scientific and Industrial Research, Dr. K. S Krishnan Road, New Delhi-110012, India.

*E-mail: husalesc@nplindia.org

## Abstract

Due to miniaturization of device dimensions, the next generation's photodetector based devices are expected to be fabricated from robust nanostructured materials. Hence there is an utmost requirement of investigating exotic optoelectronic properties of nanodevices fabricated from new novel materials and testing their performances at harsh conditions. The recent advances on 2D layered materials indicate exciting progress on broad spectral photodetection (BSP) but still there is a great demand for fabricating ultra-high performance photodetectors made from single material sensing broad electromagnetic spectrum since the detection range 325 nm - 1550 nm is not covered by the conventional Si or InGaAs photodetectors. Alternatively, $Bi_2Te_3$ is a layered material, possesses exciting optoelectronic, thermoelectric, plasmonics properties. Here we report robust photoconductivity measurements on $Bi_2Te_3$ nanosheets and nanowires demonstrating BSP from UV to NIR. The nanosheets of $Bi_2Te_3$ show the best ultra-high photoresponsivity ( ~74 A/W at 1550 nm ). Further these nanosheets when transform into nanowires using harsh FIB milling conditions exhibit about one order enhancement in the photoresponsivity without affecting the performance of the device even after 4 months of storage at ambient conditions. An ultra-high photoresponsivity and BSP indicate exciting robust nature of topological insulator based nanodevices for optoelectronic applications.

**Introduction**

Recently there is a great interest in demonstrating broadband photodetection covering from ultraviolet to infrared wavelength range especially by using novel materials like graphene, 2D dichalcogenides, topological insulators (TIs) etc. at nanodevice level aiming the future requirement of ultra-compact and high sensitive devices useful for the applications in the field of light harvesting, imaging, sensing and optical communication etc. The present commercially available photodetectors e.g. silicon which is widely studied but photoresponsivity is limited around few hundreds of mA/W. Another issue with silicon is that it doesn't show photodetection above ∼1100 nm in NIR spectrum. NIR imaging[1] is very important for many applications such as night vision, optical tomography, process monitoring etc. and due to low water absorption, wavelength range 1- 1.8 μm is more preferable and has potential commercial interest. InGaAs is the commercially available material which covers NIR spectrum but it is not sensitive to UV and cost effectiveness is also a problem. The future nano electronic devices demand compact sized high performance photosensitive materials which can cover wide electromagnetic spectrum of the light and should possess robust characteristics to various environmental conditions.

Graphene has emerged as one of the best materials for fabrication of high-performance broadband photodetector however, there are several drawbacks such as, no sizeable bandgap, limited absorption of light, fast recombination of hot carriers and low photoresposivity[2,3]. To overcome these problems alternative graphene –quantum dots, graphene -2D transistion metal dichalcogenides and graphene – topological insulator (TI) based heterostrcutures have been studied. Among these materials, TIs like $Bi_2Te_3$, $Bi_2Se_3$, $Sb_2Te_3$ are up surging as high performing broadband photodetectors. Interestingly TI based materials have bulk band gap but exhibit metallic surface states demonstrating the transport through Dirac fermions which is analogous to graphene. Thus when compared to graphene, TI has some special electronic and photonic properties that originate due to strong spin orbit interactions.

The broadband nonlinear response from visible to terahertz frequencies has been studied in TI based materials[4,5] and tuning of nonlinear response was also achieved with the help of doping in TI materials[6]. Recently MBE grown SnTe crystalline topological insulator demonstrated high photoresponsivity 3.75 A/W at 2300nm and broad spectral range was observed from visible to mid-infrared (405 nm to 3.8um)[7]. More interestingly broadband saturable absorption properties of n and p type $Bi_2Te_3$ nanoparticles have been observed at 800 and 1570 nm[6]. Bismuth telluride or selenides have attracted immense interests due to an exceptional thermoelectric[8], 3D topological insulator[9], plasmonics[10] and optoelectronic properties. $Bi_2Te_3$ is reported as a tunable plasmonic material in the visible range[10] and could be very important optoelectronic material because of small band gap which is ∼ 0.165 eV and possesses effective light absorption properties in near infrared (NIR) range. Previously Yao et. al. reported the special multifunctional photodetection property of polycrystalline $Bi_2Te_3$ film which detects intensity of incident light as well as polarization state of the incident light[11]. $Bi_2Te_3$ can be considered as a potential interesting material to study the broadband nonlinear optical and microwave properties for solid state lasers or new topological insulator based photonic devices[12,13]. The broadspectral photodetection range of $Bi_2Te_3$ is more attractive as compared to other 2D materials such as $MoS_2$, $WS_2$.The enhancement in optoelectronic properties of TI insulators by interacalation, doping etc. has been observed but robustness of TI material detecting photoconducting effects against the material deformation, impurity doping, decay due to storage in ambient conditions etc. has not been studied for $Bi_2Te_3$ material. Here first time we report the robust nature of photoconductivity in nanosheets and nanowires of $Bi_2Te_3$. These nanostructures can be used to detect the light over a wide range covering from UV, NIR and telecommunication band (1550 nm) with high photoresponsivity.

## Results

Nanosheets or flakes were deposited by using the scotch tape method and controlled FIB milling (ion beam current 50 pA and milling time < 5 Sec ) was performed on these nanosheets to obtain the nanowires of $Bi_2Te_3$ (Figure 1(a)). The energy dispersive X- ray spectroscopy (EDS) was done prior to the nanowire fabrication to know the elemental analysis and mapping of the deposited nanosheets. The inset in Figure 1(b) shows EDS spectrum detecting peaks of Bi and Te elements and atomic wt percentage was found about 39 and 61 respectively. The EDS mapping signal of Te (L$\alpha$) and Bi (M$\alpha$) is shown in the insets I and II of Figure 1(c) respectively. The HRTEM (high resolution transmission electron microscopy) characterization of nanosheets is shown in the experimental method section. The FIB based metal deposition gas injection system was used to make the Pt metal pads on the fabricated nanowires of $Bi_2Te_3$. The schematics used for optoeletronic characterization of the nanowire devices is shown in the Figure 1(d) where light was illuminated uniformly covering the whole device.

The broadspectral photodetection was first studied in $Bi_2Te_3$ nanosheets and the false colour FESEM image of fabricated device with two platinum contacts is shown in the inset of Figure 2(a). The device was illuminated with UV, visible, NIR lights and time dependent rise or decrease in device's current was measured. The sharp sudden rise or decrease in current was observed when the incident light was either switched ON/OFF respectively. Different bias voltages were applied and the corresponding increase in photocurrent was measured which is shown in Fig. 2(a). The increase in current depends on the applied bias voltage through the relation, $I_g = 2I_p e\mu(W/L)V_{sd}$ where $\mu$ is mobility, $e$ is the electronic charge, $W$ is the width and $L$ is the channel length of the device. With higher applied bias voltage, higher photocurrent was observed. The obtained data clearly shows that $Bi_2Te_3$ nanosheet exhibits broad spectral photosensitivity. The sensitivity towards 1550 nm light is very important since it is widely used in telecommunication applications. The rise and decay times were obtained by fitting the curves with equations $\boldsymbol{I} = \boldsymbol{I_0}(\boldsymbol{1} - \boldsymbol{e}^{\frac{-t}{\tau_r}})$ and $\boldsymbol{I} = \boldsymbol{I_0}(\boldsymbol{e}^{\frac{-t}{\tau_d}})$ respectively where $\boldsymbol{\tau_r}$ and $\boldsymbol{\tau_d}$ are the rise and the decay time constants respectively (supplementary information Figure S1). The $\boldsymbol{\tau_r}$ and $\boldsymbol{\tau_d}$ were observed in millisecond time scale for UV, visible and NIR lights illuminations (Table 1). The characteristic values obtained in the present study are comparable and/or better with results reported on topological insulator based nanowires or films. The curves shown in Fig. 2(a) clearly demonstrate the $Bi_2Te_3$ nanosheet shows broad spectral sensitivity and device was found stable at different bias voltages (25 mV – 300 mV). The responsivity is a very essential and an important parameter for any kind of photodetector and can be estimated through the measured photocurrent, wavelength of light, device area and power density of the incident laser light. The Fig. 2(b) shows the responsivity curves of the nanosheet as a function of applied bias voltage. For UV, visible and NIR light irradiations, we observed linear increase in the responsivity values. The responsivity was found better for NIR light compared to visible and UV lights. The photocurrent and responsivity of the nanosheet device were further characterized as a function of laser power density (NIR 1550nm) and the curves are shown in the Figure 2(c). As expected, the decrease in responsivity was observed as a function of increase in the laser power density which is consistent with the other published results on topological insulators based nanowires[14], films[15]. The curves represent that the nanosheets respond well to incoming light from UV to NIR region and this material could be the alternative choice for the present photodetectors either to replace silicon which do not have NIR sensitivity beyond ~ 1100 nm or InGaAs which do not show UV sensitivity. Three more nanosheets were tested further where we find similar broad spectral photoresponse and responsivity. The device images and photoconductivity measurements for these nanosheets are shown in the supplementary information Figure S2-S4.

Photodetectors based on nanowires are often show ultrahigh performance properties compared to film or bulk counterparts. Earlier reports on fabricated nanowires of $Bi_2Se_3$ showed high performance properties but was not investigated for telecom wavelength[14]. $Bi_2Te_3$ with bandgap $\sim$ 0.165 eV[9] is more appealing material for NIR sensing applications and we fabricated nanowires by FIB milling method ( NW1 and NW2 as shown in insets of Figure 3(a) and 3(d) respectively). The fabricated NW1 was first tested for UV and NIR light irradiations and time dependent changes in photocurrent due to light ON/OFF cycles were monitored at a constant bias voltage and at different laser power densities as shown in the Fig. 3(a). After each cycle, the power density was slightly increased and the corresponding rise in photocurrent was monitored. Thus the increase in the amplitude of photocurrents indicates corresponding increase in the power density of laser light illumination in $mW/cm^2$ as shown by the numbers. For every increase in the power density of the laser at a fixed bias voltage shows the increase in the photocurrent and higher photocurrent was observed for the 1550 nm. The periodic On/OFF cycles with different power densities and laser wavelengths represent the stability of the device working at the room temperature. The device was also exposed under visible laser radiations 532 nm and the bias voltage dependent photocurrent increase was monitored which is shown in the supplementary information Figure S5. The photocurrent values measured at 300 mV bias voltages were used to know the responsivity and detectivity curves for UV, visible and NIR wavelengths are shown in the Fig.3 (b). The higher responsivity was observed for the NIR light (1550 nm). The laser power dependent photocurrent measurements were further repeated at different bias voltages and corresponding curves for NIR light (1064 nm) illumination are shown in the Figure 3(c). The fabricated nanowires were found nicely detecting the incident light at different bias voltages and the linear increase in photocurrent was observed as a function of incident power density. The bias voltages were applied from 25– 300 mV and responsivity curve for the data at 300 mV is shown in the inset of Fig 3(c). At higher power density, generation of more electron –hole pair is expected which contribute to the enhancement in photocurrent and the linear response suggest that traps or deformations present in the fabricated material do not affect the photoconductivity.

The photoconductivity measurements were repeated for another fabricated $Bi_2Te_3$ nanowire device (NW2, inset Figure 3(d)). The time dependent photocurrent measurements were performed under UV, visible and NIR wavelengths and responsivity curves as a function of applied bias voltage are shown in Fig. 3(d). The time dependent photocurrent curves for UV, Visible wavelengths at different bias voltages are shown in the Figure 3(e) and for NIR wavelength is shown in the supplementary information Figure S6. The sudden rise or drop in photocurrent measurement is due to switch ON or OFF cycles of the laser light exposure during the constant applied bias voltage. The reproducibility in broad spectral photodetection was clearly observed. Figure 3(f) represents the performance of NW2 device under the illumination of NIR laser 1550 nm at 300mV bias voltage. The maximum photoresponsivity about∼ 778 A/W was observed at lower power density and the inset shows the detectivity of the NW2 device as a function of incident laser power density. The values of responsivity and detectivity at 300 mV bias voltage are shown in Table (1) which are quite competitive compared to other reported values. The reproducibility of broadspectral photoresponse was also observed in NW3 device and the data is shown in the supplementary information Figure S3.

Topological insulators based materials possess intrinsic robust transport properties and expected that presence of nonmagnetic impurities and material deformations do not affect them. Here we used FIB fabrication process to test the robust photoconducting nature of $Bi_2Te_3$ nanosheet and nanowires because FIB milling technique inherently implants Ga ions and some deformation in nanosheet during fabrication of nanowires from the nanosheet. First we have studied photoresponse of deposited nanosheet contacted with Pt metal electrodes which is shown in the inset of Figure 4(a). The time dependent photocurrent measurements under the illumination of UV, Visible and NIR (Figure S7) lights were first carried out to confirm the broad spectral optoelectronic properties. In the second step nanosheet was transformed into a nanowire form Fab_NW-1 (inset of Figure 4(b)) and the similar to nanosheet, broad spectral photoconductivity measurements were repeated and shown in the Fig. 4(b). The fabricated nanowire clearly showed the broad spectral nature of photoresponse. The Fab_NW-1 wire was further narrowed in second step to get the Fab_NW-2 device ( inset Figure 4(c)) and the nature of broad spectral photoconductivity was clearly noticeable (curves in Fig. 4(c). In $3^{rd}$ step Fab_NW-2 device was further narrowed down to ∼ 160 nm where we expected more distortions and Ga implantations (device image in inset Figure 4(d)). The time and applied bias voltage dependent photoconductivity measurements for this device are shown in the Fig. 4(d). Still the broadspectral photoresponse is clearly visible. The comparative bar chart of photoresponsivity observed at 300 mV bias voltage starting from the nanosheet to milled nanowire device Fab_NW-3, is shown in the Figure 4(e). The bars clearly represent that there is no degradation of the photocurrent even after the milling was performed 3 times. The slight rise in photocurrent for all wavelengths was observed. The better performance in photodetection could be due to the nano confinement effects and more electron hole pair generation for every effective incident photon. The reproducibility of robust nature photoconductivity under repetitive milling operations was also checked on other device and data is shown in the supplementary information Figure S3.

The aging parameters such as the functioning of the device over a longer period, storage of nanowire device at ambient conditions are important to know the robustness of device. Important to note that nanowire based devices could be damaged due to moisture, exposure to ambient conditions and longer duration storage and hence device's performance may be degraded over the time. In Figure 4(f), we have tested the robustness of the device over 4 month's time. The curves in the figure show time dependent photocurrent measurements at constant bias 300 mV. The black curve shows the measurements done after the fabrication of the device within a day time whereas the green and pink curves represent the photoconductivity measurements performed after 35 and 120 days interval of time, respectively. The Figure S8 shows the changes in the gain values and a moderate degradation in device's performance was observed. Note that the device was stored at ambient conditions and some degradation is expected. Overall still it was convincingly showing the broadspectral photoresponse and photoresponsivity of about ~210 A/W at 1550 nm was observed. Previously a very slow growth of oxide layer, ~ 2nm in 5700 hrs was observed in $Bi_2Te_3$ material under the continuous exposure of air[16]. The oxidation of $Bi_2Te_3$ shifts the Fermi level towards up or down direction with respect to the Dirac point and this could be the reason behind slight degradation of photoconductivity in our devices. This indicates robust performance of the device and was not much affected by the ambient conditions. The detectivity and photoconductive gain are important parameters for the imaging applications and can be estimated from the following relations $D = (R\sqrt{A}) \; / \; (2eI_d)^{1/2}$ and $G = R \times \frac{hc}{\lambda e}$ respectively, where $R$ is the responsivity, e is electronic charge, $I_d$ *is* dark current, $A$ is the active area, $h$ is Planck's constant, c is the velocity of light. The detectivity and photoconductive gain values are reported in the Table (1).

## Discussion:

Nanoplates of $Bi_2Te_3$ represent a new class of tunable plasmonic material[10] and these devices can be further used to study the light matter interactions in various heterostructures. The previous reports on nanostructure based devices showed high performance optoelectronic properties due to quantum confinement effects[14,17,18]. The $Bi_2Te_3$ nanowires fabricated and studied here show higher photoresponsivity ( ~ 778 A/W at 1550nm and 300 mV bias) which is better than the earlier reported responsivity ( 35 A/W at 532 nm) obtained from photodetectors based on graphene –$Bi_2Te_3$ heterostructures[19] and MBE grown $Bi_2Te_3$ films[20]. The higher values of the responsivity observed in our devices could be due to the strong light matter interactions, high quality of the deposited material, the large surface area to volume ratio favored by nano milling and the carrier transport through efficient surface states. This is because topological insulator based materials show transport properties through topological surface states (TSS) and bulk is insulating in nature, the efficient carrier transport property of TSS may be favourable to achieve the high performance photodetector. Note that the fabricated $Bi_2Te_3$ nanowire devices are better due to increase in surface to volume ratio, abundant surface carriers and compared to single layer dichalcogenides devices, absorption of light could be better since their thicknesses are large compared to single or bilayer devices. Taking into an account the short penetration length of the incident light, the top few layers of the material contribute more to the photocurrent generation hence in case of nanosheets or nanowires of TIs, more light absorption is favourable.

The significant and reproducible photocurrent was detected in all the devices studied here under periodic On/Off cycles of the laser light, the sudden rise and decay in photocurrent correspond to the On and Off state of the incident laser light. The saturation in photocurrent is clearly visible and the faster decay indicates that there are not so much charge trapping centers after the fabrication process got complete. Further, material inhomogentiy formed due to Ga implantation is not much affecting the photoresponse of the material when compared it with the bare nanosheet. The rise or decay times in sec or ms have been observed in many 2D material based systems and topological insulator based systems [21,22,23] [19,24]. Note that the rise or decay time constant values are estimated here either using the exponential rise or decay fit equations. However we find that rise time or decay time constant in ms are in good agreement (Table 1) with the published literature and show competitiveness for photodetector applications and the fast response indicate the efficient carrier transport facilitated may be due to the robust nature of the TSS and high mobility[25,26] which is far better as compared with the transition metal dichalcogenides[27,28]. The broad spectral photodetection based on nanobelts, nanoribbons, nanosheets and emerging 2D materials have been recently published in a review article[29]. The nanobelts of molybdenum trioxide[30], nanosheets of InSe ($\lambda$ = 850 nm)[31] and $SnS_2$ ($\lambda$ = 850 nm )[32] show responsivity values 56 A/W, $2.975 \times 10^3$ and $1.22 \times 10^{-8}$ A/W respectively (2,3,4). Other 2D materials like layered black phosphorus (BP)[33], BP/ monolayer $MoS_2$[34], ferroelectric polymer film gated with ferroelectric material[34], graphene $Bi_2Te_3$[18], $Bi_2Te_3$-SnSe-$Bi_2Te_3$[35], Bi/$WS_2$/Si[36] and $\alpha$-$In_2Te_3$[37] show responsivity values $0.4 \times 10^{-3}$,3.54, 2570, 35, 5.5, 0.42 and 44 A/W, respectively (5,6,7,8,9,10,11). Compared to these materials, responsivity obtained with nanowires of $Bi_2Te_3$ is much more competitive taking into account of single material and no gate field effects.

Robust photoconduction observed here indicate that efficient carrier transport through TSS is resistant to material deformations caused by ion milling and non magnetic Ga impurities. $Bi_2Te_3$ in combination with other material such as $WS_2$ and Si have been demonstrated the broad spectral photoresponsivity[21,38]. The nanoflakes of $Sb_2SeTe_2$ and thin films of $Sb_2Te_3$ have also shown high performing visible[39] and near infrared [24] photodetection recently. Huang et al observed that transport through TSS can tolerate the surface oxidation and molecules absorbed on the samples surface[40] and our previous work also showed the robustness of TSS towards the Ga ion milling and inherent material deformations[26,41].

In Fig. 4(f) we observed that even at sequential milling operations, the photosensitivity do not degrade. Here we explain that the observed increase in photocurrent mostly coming from surface states by considering the milling of the device area i.e. changes in the nanowire widths corresponding to the change in the photocurrent. Figure 5 shows the energy dispersion diagram illustrating the various possibilities of optical excitation and photocurrent generation. The Fig. 5 shows that under illumination of light, hole – electron pairs are generated due to the possibility in both bulk (I) as well as in TSS (III). Since the topological insulator materials show metallic surface states having Dirac cone distribution hence the light can be detected over a broad spectral range similar to that of graphene [5]. On the other hand, the bulk band gap of $Bi_2Te_3$ is ~ 0.165 eV[9] and absorption of light will eventually create the electron hole pairs.

The total conductance is the combination of conductance coming from the surface states and contribution of bulk conductivity and can be written as $G_{total} = G_{Surface} + G_{Bulk}(WH/L)$ , where $W$, $H$ and $L$ are width, height and length of the bulk channel. The defects, implanted Ga ion may help to form vacancies and antisite defects which could dope the film that may shift the Fermi level either to the conductance or valence band resulting the conductance dominated by the bulk channel at room temp. Note that topological insulators systems are predicated to be robust against any nonmagnetic impurities or deformations. Previously we have studied low temp transport studies and found quantum oscillations at low temp demonstrating the robust nature of TSS[26,41]. The absorption of light and generations of electron hole pairs are more effective in few layers of $Bi_2Te_3$. The bulk $Bi_2Te_3$ material shows large mean free path length ~ 60 nm[42] that makes easy for the excited carriers to likely drifting towards the conducting surface state channel (step II) as shown in the Fig. 5. The femtosecond ultrafast spectroscopy and angle resolved photoemission spectroscopy revealed that optically excited carriers, accumulated in bulk conduction (metastable population) feeds a no equilibrium population of the surface states[43]. Under bias condition, these accumulated carriers contributed to the rise in the photocurrent. The fabrication approach indicates the narrowing of the sample i.e. contribution from the bulk channel is less but even though the photocurrent found increasing which suggests that contribution of bulk is less if we compare the nanosheet to Fab_NW-3 dimensions (inset Fig. 4(d)). The reason could be a stronger surface contribution to the conductance, increased surface to volume ratio, enhanced free carriers generation by quantum confinement effects etc.

The optical absorption of TI based materials is strongly depends on thickness of the material and dramatic enhancement in the performance of the photodetector was theoretically seen when thickness of the material was reduced to several quintuple layers[5]. Theoretically the optical conductivity of topological insulator $Bi_2Se_3$ thin films was seen over a broadspectral region spanning from infrared to visible[44]. The different routes of optical absorption in thin films of TI were observed and transitions were mainly arising from intraband, interbands, and surface states in the valence band to surface states in conduction bands. Previously the MBE grown $Bi_2Te_3$ thin films on Si were used to detect photoresponses at NIR wavelengths and the responsivity was observed ~ 3. 64x $10^{-3}$ and 3. 32x $10^{-2}$ A/W for 1064 and 1550 nm wavelengths respectively[20]. Here we have observed ultrahigh responsivity of ~ 778 A/W at NIR excitation with 1550 nm and the overall performance values of $Bi_2Te_3$ nanosheets and nanowires are shown in table 1, which are either better or competitive. This indicates that nanosheets or nanowires of $Bi_2Te_3$ material have potential technological use in photodetection without the need of high drain to source bias or gate voltages (Table 1).

## Conclusion:

The nanosheet and/or nanowires of $Bi_2Te_3$ show high performance optoelectronic properties. The clear broad spectral photodetection for UV, visible and NIR wavelengths was observed in all the measured samples. The robust nature of photoconductivity in $Bi_2Te_3$ against the nano milling conditions is evident. The devices stored at ambient conditions show slight degradation in optoelectronic properties while keeping the broad spectral response unaffected. The observed robust ultra-high photoresponsivity and broadband photodetection indicate exciting optoelectronic applications which can be exploited further as robust photodetectors are having potential applications in nanocircuits, nanodevices, photodetectors and sensors.

## Method:

**High resolution transmission electron microscopy** (HRTEM, model: Tecnai G2F30 STWIN) was employed to characterize the specimens of $Bi_2Te_3$. In general, nano-sheets of $Bi_2Te_3$ normally with either elongated or flat morphologies were delineated throughout in the microstructure (Figures. 6(a) and (c)). The elongated small nanosheets appeared very tiny with the length and width of about 50 and 5 nm, respectively (Figure. 6(a)). A corresponding selected area electron diffraction pattern (SAED) from the bundles of elongated morphologies (Figure. 6(a)) exhibits a set of Debye rings (inset in Figure. 6(a)). The important planes of $Bi_2Te_3$ rhombohedral crystal structure (lattice constants: a = 1.045 nm, $\alpha$ = 24.13°, space group: $R\bar{3}m$, reference: JCPDS card no. 850439) with interplanar (d) spacings 0.32, 0.24 and 0.22 nm corresponding to hkl: 221, 433, and $1\bar{1}0$, are marled as 1, 2, and 3, respectively on the inset of Figure 6(a). A high resolution image of atomic planes from the encircled region in Figure. 6(a) further shows the atomic planes (Figure 6(b)) with the interlayer separation of about 0.32 nm corresponding to hkl indices of 221 of $Bi_2Te_3$ rhombohedral crystal. In another flat morphology of nano-sheets as depicted in Figure 6(c), the overall area is larger with respect to elongated morphology (Figure. 6(a)). A corresponding atomic scale image recorded from the encircled area of Figure. 6(c) exhibits a set of atomic planes of d values of about 0.24 nm with hkl indices of 433 of $Bi_2Te_3$ rhombohedral crystal.

## Acknowledgments

A.S. acknowledges the SRF fellowship of Council of Scientific and Industrial Research, India. S.H. and A.S. acknowledge CSIR's Network project "Aquarius" for the financial support. We sincerely thank HOD DU 2, Dr. V N Ojha for his support and encouragement.

## Author contributions

A.S. deposited and localized the nanosheets, performed metallization using sputtering system and carried out all the optoelectronic measurements. A.S analysed all the the data and fabricated the nanodevices with SH. A.K.S. performed and analysed the HRTEM data. T.D.S. provided FIB tools, operational support and materials. S.H. conceived and supervised the research and wrote the manuscript.

## Additional information

## Competing financial interests

The authors declare no competing financial interests.

## Corresponding Author

*E-mail: husalesc@nplindia.org

## Figures

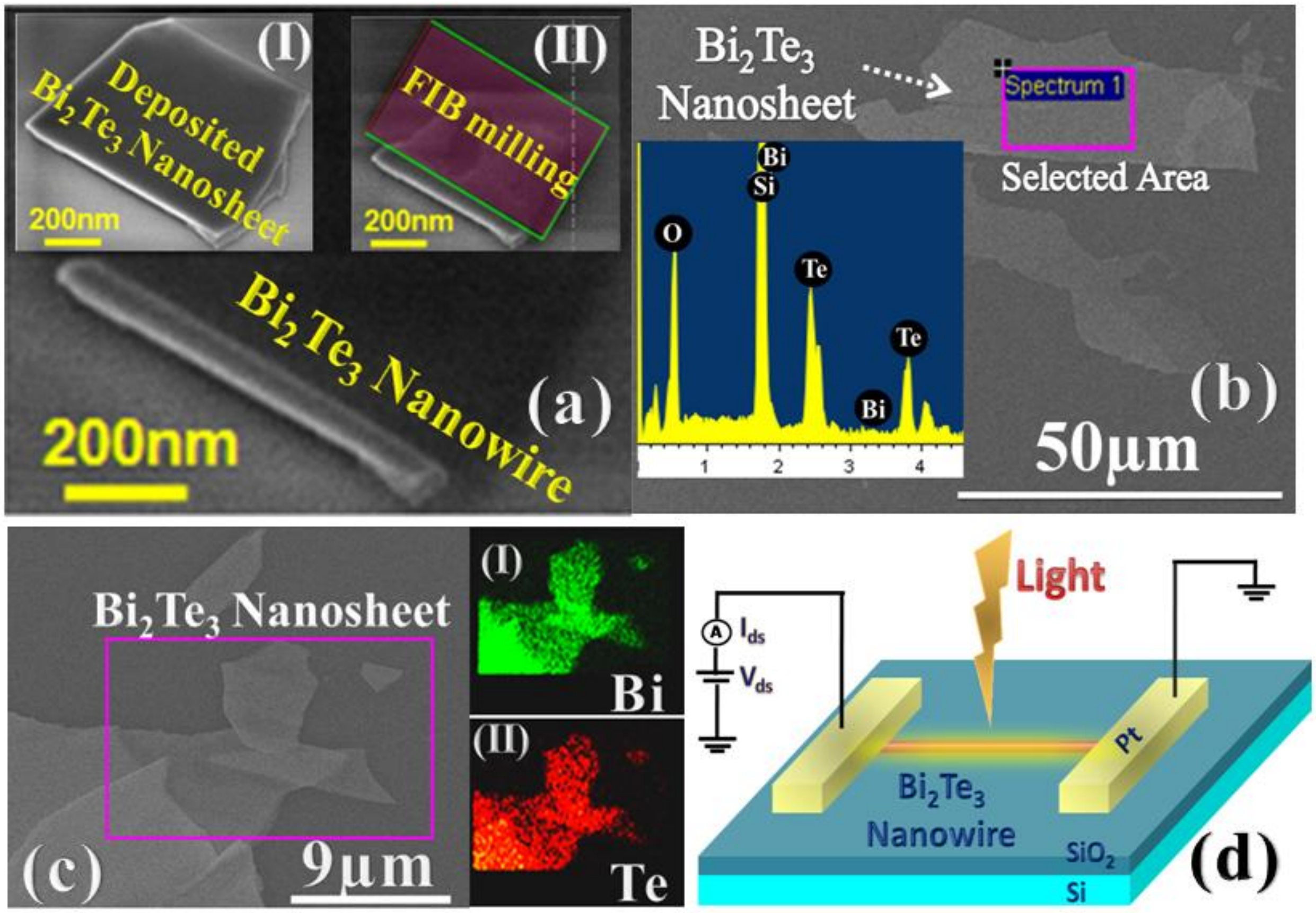

**Figure 1.** Nanosheet and nanowires of $Bi_2Te_3$. Inset I and II in Fig. (a) show the $Bi_2Te_3$ nanosheet deposited using scotch tape method and selective FIB milling of the nanosheet. The Fig. (a) shows the FIB fabricated $Bi_2Te_3$ nanowire. Fig. (b) shows the FESEM image of $Bi_2Te_3$ nanosheet used for elemental analysis and inset represents the EDS characterization of the selected portion of the nanosheet. The rectangle in Fig. (c) shows the area used for EDS elemental mapping for Bi and Te as shown in inset I and II respectively. Fig. (d) represents the schematics used for the optoelectronics characterization.

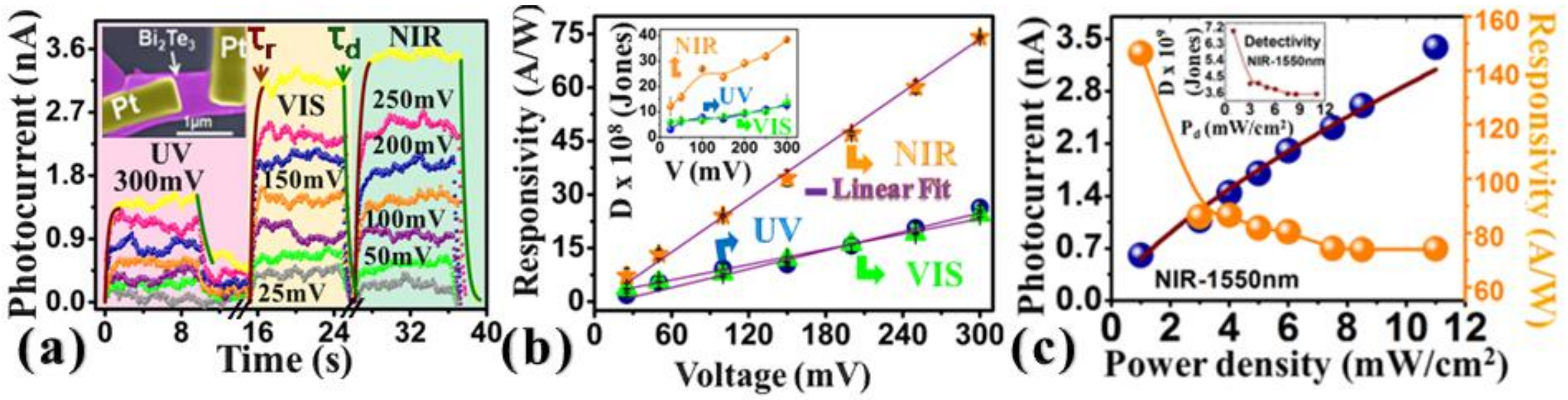

**Figure 2** Optoelectronic characterizations of as deposited $Bi_2Te_3$ nanosheet. Time dependent photocurrent measurements at different bias voltages and incident wavelengths (Fig. (a)). Inset is the false colour FESEM image of the $Bi_2Te_3$ nanosheet device. Fig. (b) shows the bias voltage dependent responsivity curves measured under different light irradiations.  Fig. (c) represents the NIR (1550 nm) laser power density dependent photocurrent and responsivity curves. Inset in Fig. (c) shows the laser power density dependent detectivity curve.

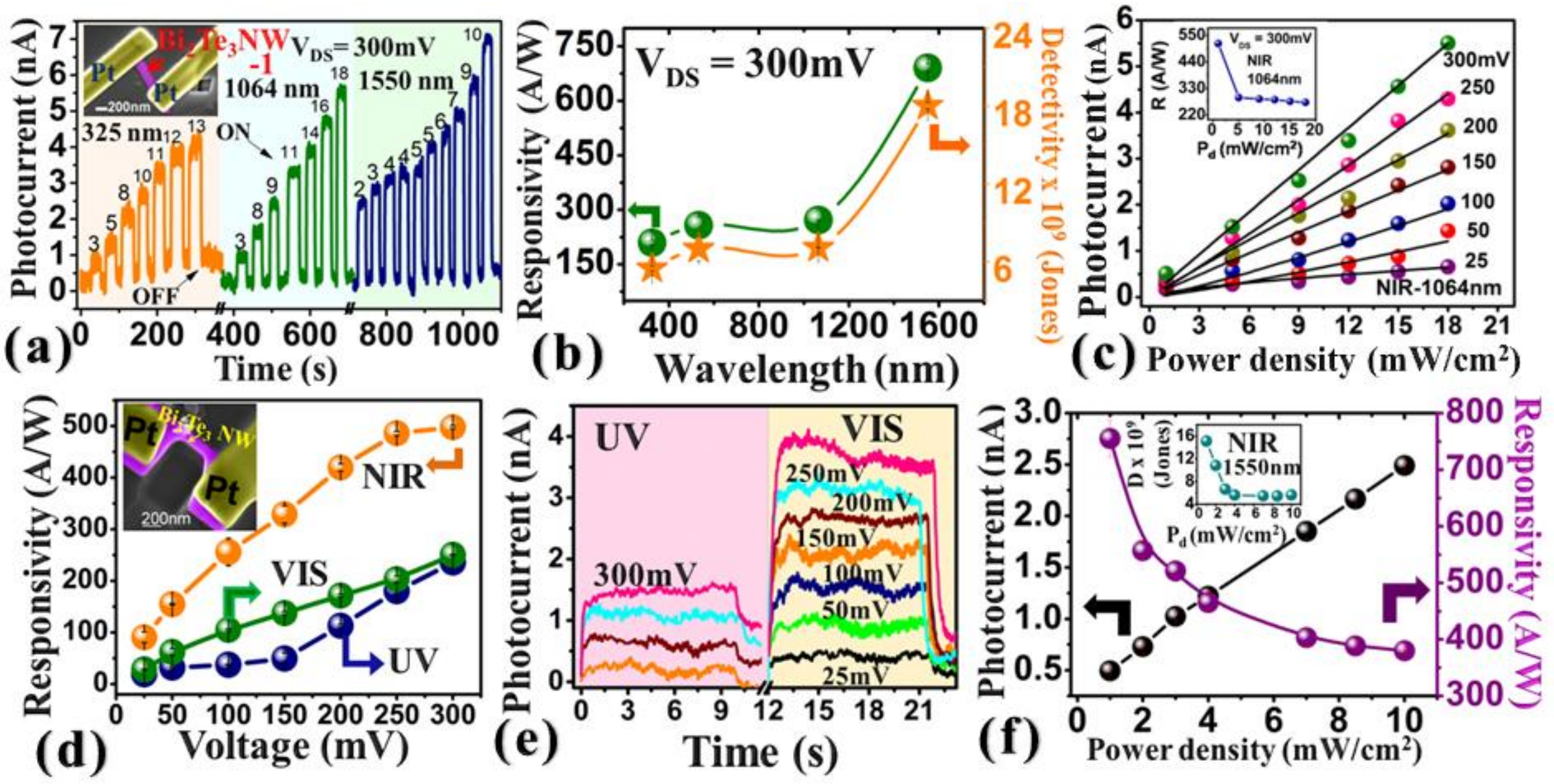

**Figure 3.** Optoelectronic characterization of FIB fabricated $Bi_2Te_3$ nanowires. Inset in Fig. (a) represents the $Bi_2Te_3$ nanowire device (NW1) used for time and incident power dependent photocurrent measurements under UV and NIR light illuminations. The responsivity and detectivity curves obtained for UV, visible and NIR lights are shown in Fig. (b). The photocurrent dependence on incident power density (Fig. (c)) at different bias voltages and inset shows the responsivity curve. Fig. (d –f) represent the optoelectronic characterization for NW2 device (inset Fig. (d)). Fig. (d) shows the bias voltage dependent reponsivity curves observed for different wavelengths. The photocurrent measurement as a function of time in presence of visible and UV light illuminations at different bias voltages is shown in Fig. (e). Fig. 3(f) shows NIR laser power density dependent photocurrent and responsivity curves. Inset represents the power density dependent detectivity curve.

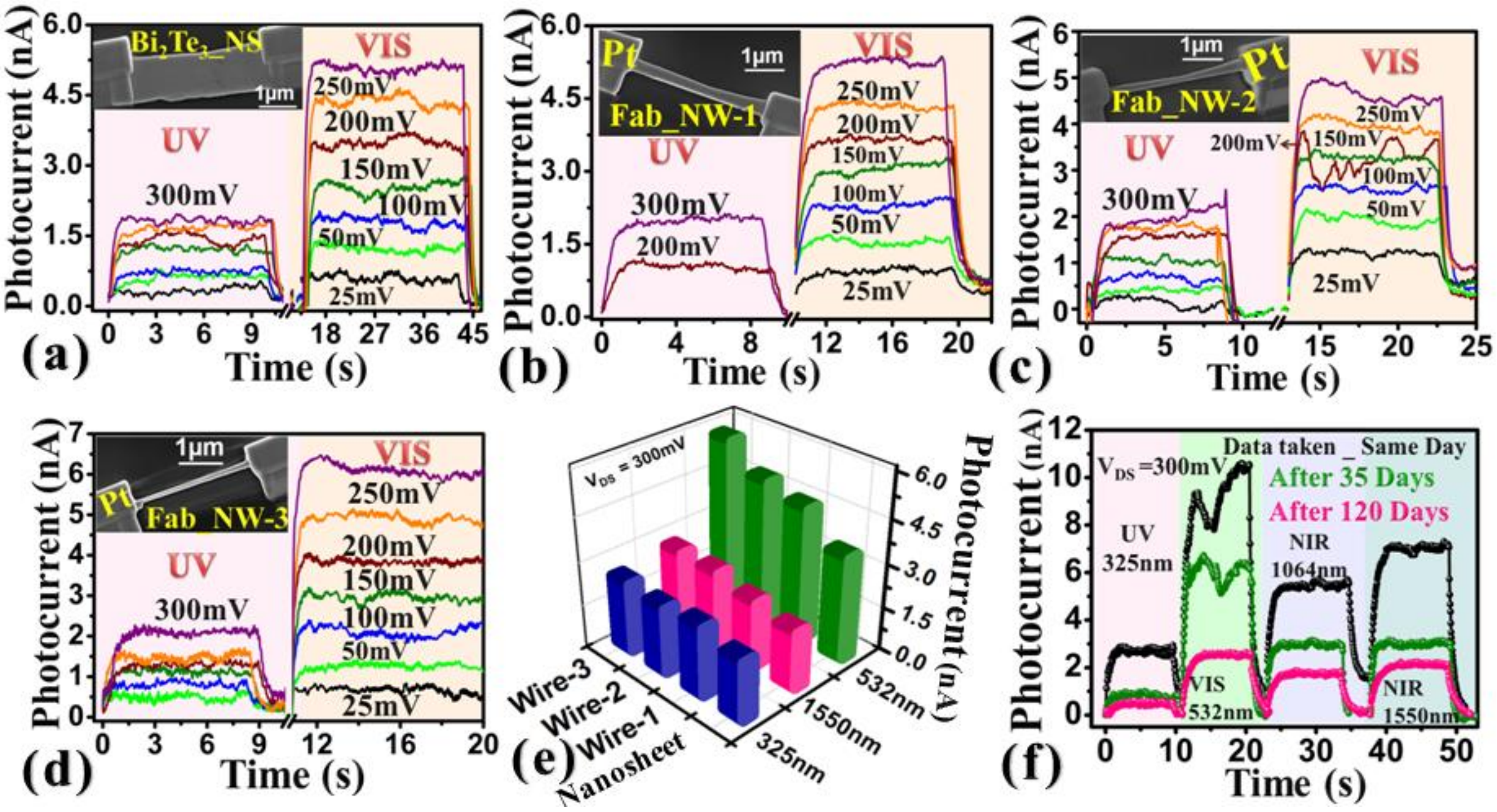

**Figure 4.** Robustness and enhancements in photoconduction measurements. Time, applied bias and illumination light dependent photocurrent measurements for as deposited nanosheet (Fig. (a)), Fab_NW1 (Fig. (b)), Fab_NW2 (Fig. (c)) and Fab_NW3 (Fig. (d)). Device images are shown in the insets of the respective graphs. Fig. (e) represents the photocurrent comparison between nanosheet and fabricated nanodevices. Fig. (f) represents photocurrent response curves measured after ambient storage conditions as shown in the graph.

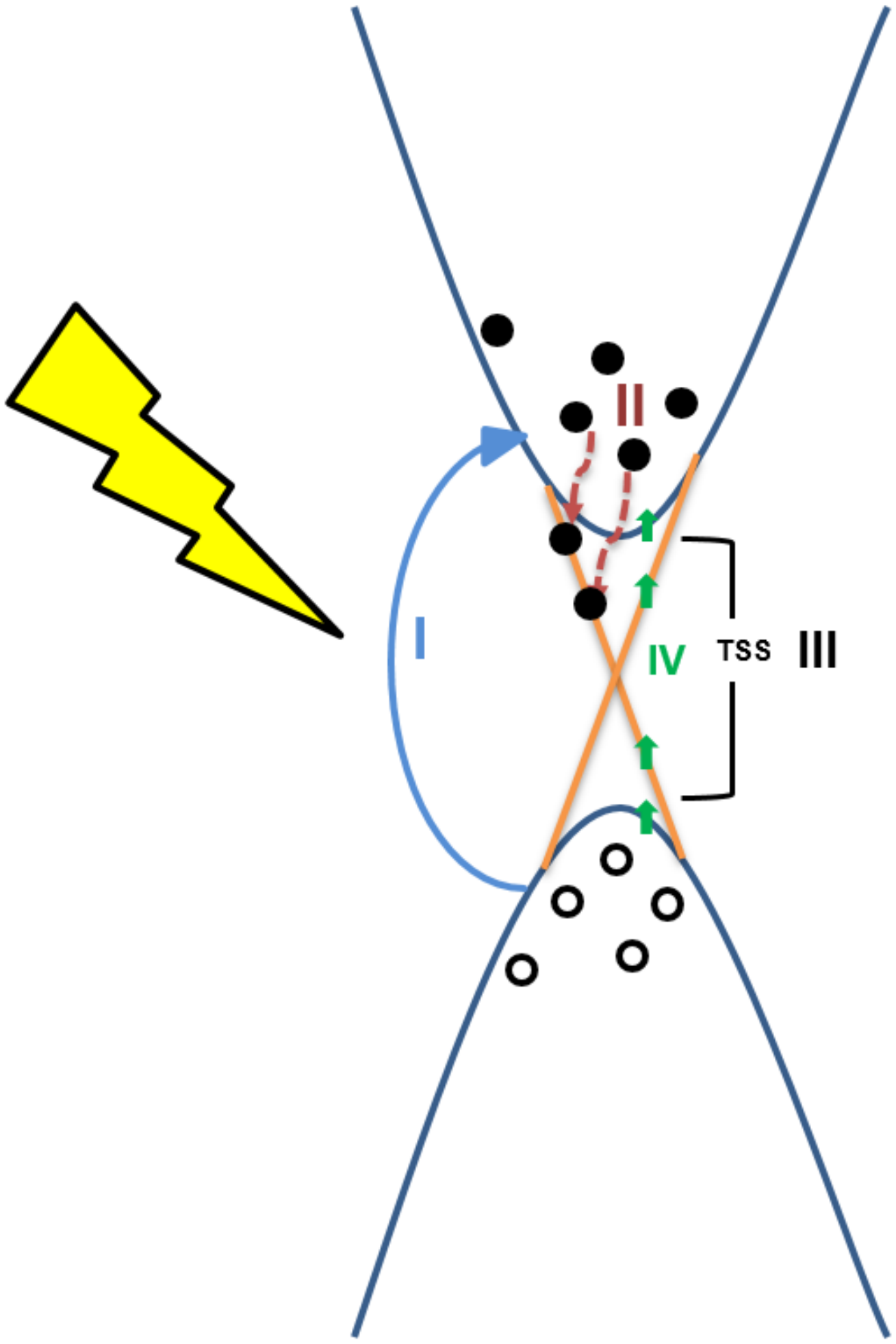

**Figure 5** Schematic representation of energy band diagram illustrating the various possibilities of optical excitation and photocurrent generation

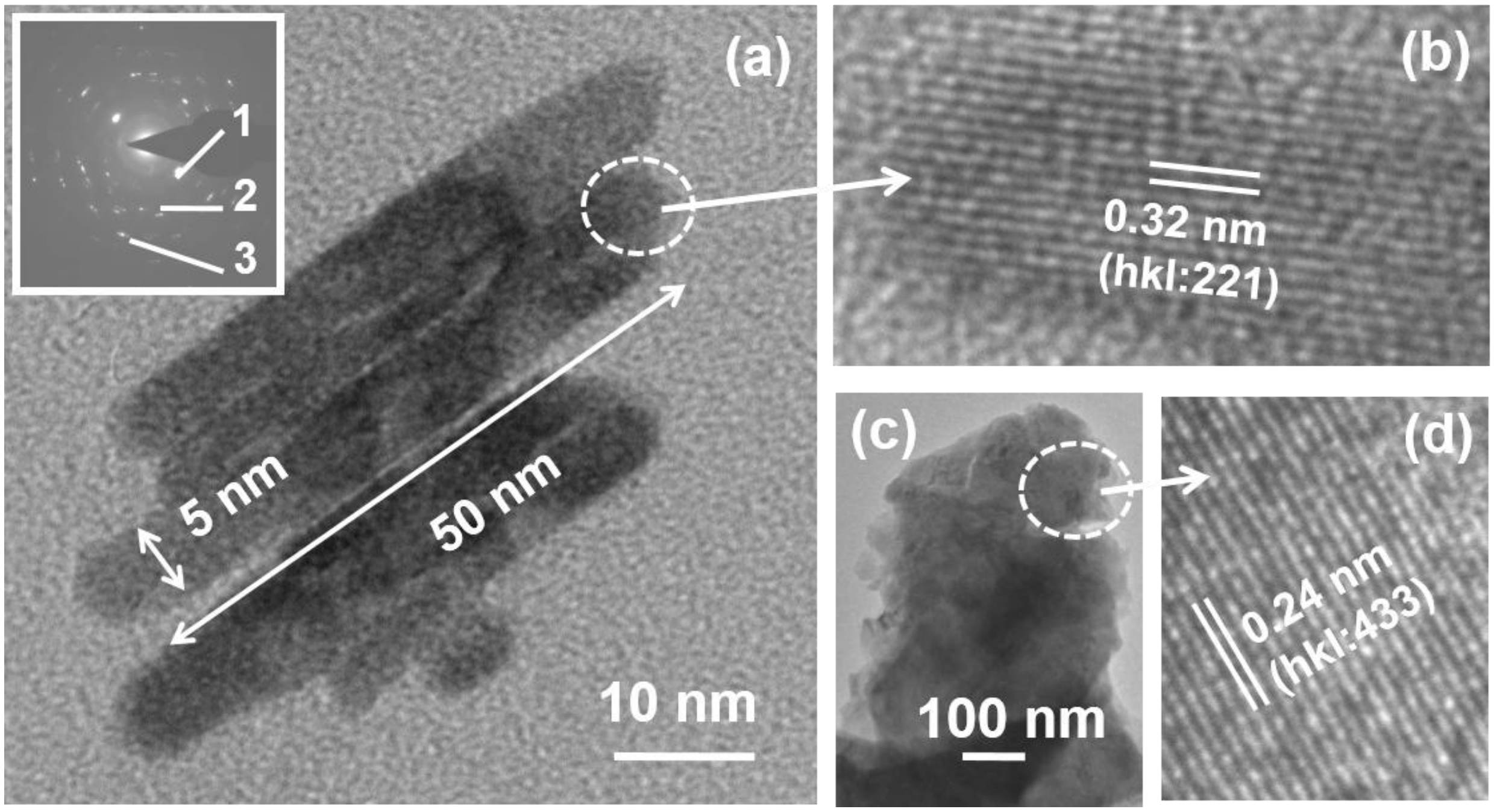

**Figure 6.** HRTEM results showing, (a) bright field micrograph of elongated morphology (tiny nanosheets), (b) corresponding atomic scale image, (c) bright field micrograph of another large nanosheet, (d) corresponding atomic scale image. Inset in (a) an electron diffraction pattern.

**Table 1.**

| **Material** | **λ (nm)** | **R ($AW^{-1}$)** | **D (Jones)** | **Gain/EQE** | **Rise/Decay $\tau_r$ / $\tau_d$, (s)** |
|---|---|---|---|---|---|
| $Bi_2Se_3$ nanosheets[45] | ------- | 20.48 x $10^{-3}$ | --------- | 8.36 | 0.7/1.48 |
| $Sb_2Te_3$ film[24] | 980 | 21.7 | 1.22 x $10^{11}$ | 27.4 | 238.7 / 203.5 |
| Bi film[46] | 370 | 250 x $10^{-3}$ | ---------- | ---------- | 0.9/1.9 |
| $Bi_2Te_3$ film / Si[47] | 808 | 924.2 | 2.38 x $10^{12}$ | 1421 | 0.045/0.047 |
| $Bi_2Te_3$-Graphene[19] | 532 | 35 | ---------- | 83 | 8.7x$10^{-3}$/ 14.8x$10^{-3}$ |
| | 980 | 10 | | 11 | |
| $Bi_2Se_3$ nanowire (NW)[14] | 1064 | 300 | 7.5 x $10^9$ | 350 | 0.550/0.400 |
| $Bi_2Se_3$ (NW)/ Si[15] | 808 | 24.28 | 4.39 x$10^{12}$ | 37.4 | 2.5 x $10^{-6}$ / 5.5 x $10^{-6}$ |
| Polycrystalline $Bi_2Te_3$ / Si [21] | 635 | 1 | 2.5 x $10^{11}$ | ----------- | 0.1/0.1 |
| $WS_2$ -$Bi_2Te_3$ [48] | 370-1550 | 30.4 | 2.3 x $10^{11}$ | --------- | 0.020 / 0.020 |
| **$Bi_2Te_3$ Flake** | **325** | **26.82±0.33** | **1.29 x $10^9$** | **102±0.46** | **0.28 / 1.6** |
| | **532** | **24.72±0.17** | **1.5 x $10^9$** | **57.72±0.28** | **0.37/0.42** |
| | **1550** | **74.32 ± 4** | **3.8 x $10^9$** | **59.56±0.56** | **0.42/0.44** |
| | | | | | |
| **$Bi_2Te_3$ NW (This work)** | **325** | **238 ± 0.73** | **4 x $10^9$** | **909.61±0.2** | **0.43/ 0.95** |
| | **532** | **251 ±0.32** | **4.5 x $10^9$** | **586.15±0.1** | **0.48/0.54** |
| | **1550** | **778 ± 0.05** | **1.2 x $10^9$** | **623.59±0.2** | **0.50/0.60** |

**Table (1)**. Topological insulator material based photodetectors and performance comparison with our results.

## Robust broad spectral photodetection (UV-NIR) and ultra high responsivity investigated in nanosheets and nanowires of $Bi_2Te_3$ under harsh nano-milling conditions

Alka Sharma[1,2], A. K. Srivastava[1,2], T. D. Senguttuvan[1,2] and Sudhir Husale[1,2]*

[1]Academy of Scientific and Innovative Research (AcSIR), National Physical Laboratory, Council of Scientific and Industrial Research, Dr. K. S Krishnan Road, New Delhi-110012, India.

[2] National Physical Laboratory, Council of Scientific and Industrial Research, Dr. K. S Krishnan Road, New Delhi-110012, India.

*E-mail: husalesc@nplindia.org

## Supplementary Information contents:

1. Rise and Decay time fit

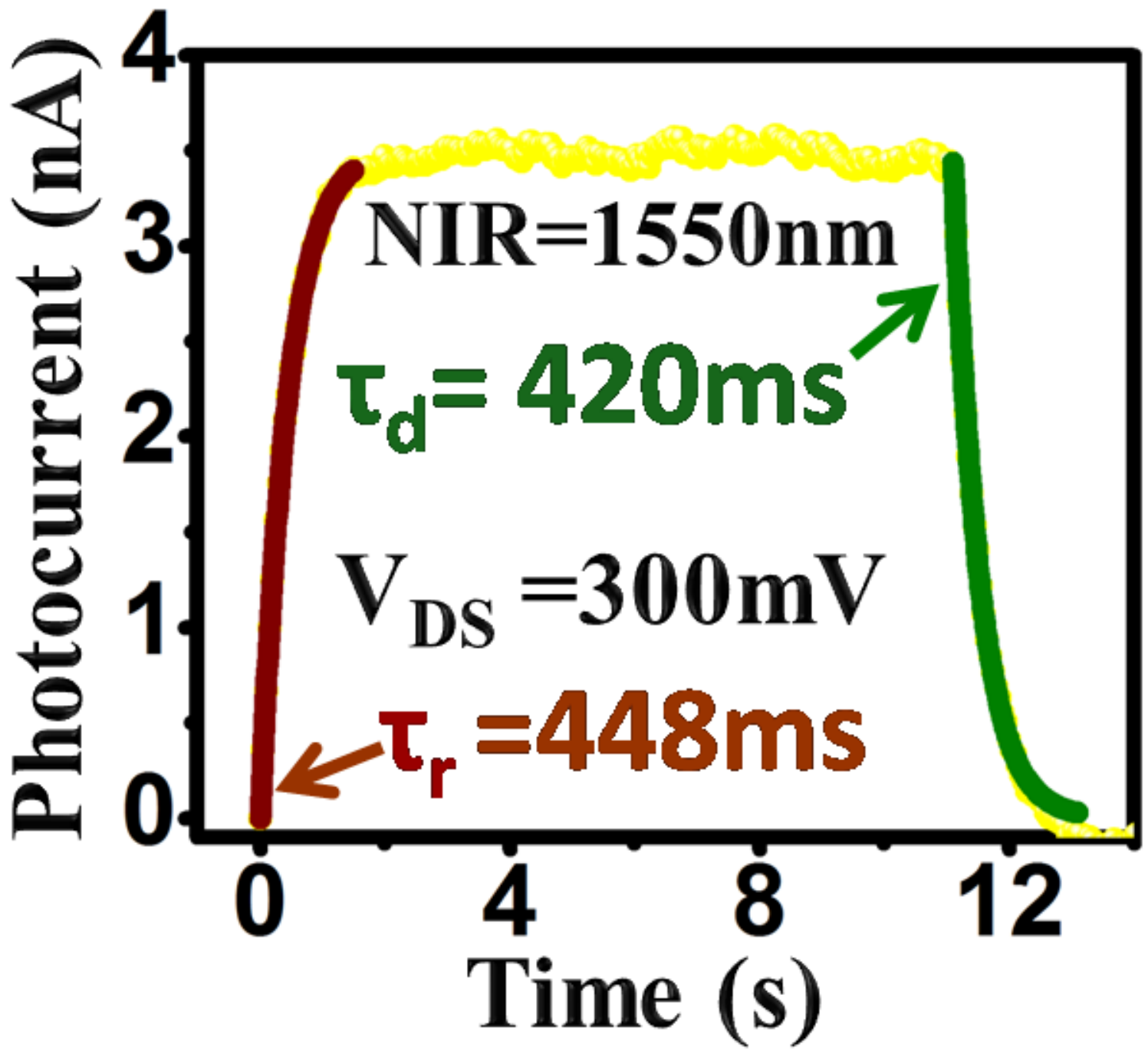

**Figure S1:** Rise ( red ) and decay (green) curve fitting.

## 2. Optoelectronic characterization of $Bi_2Te_3$ nanosheets

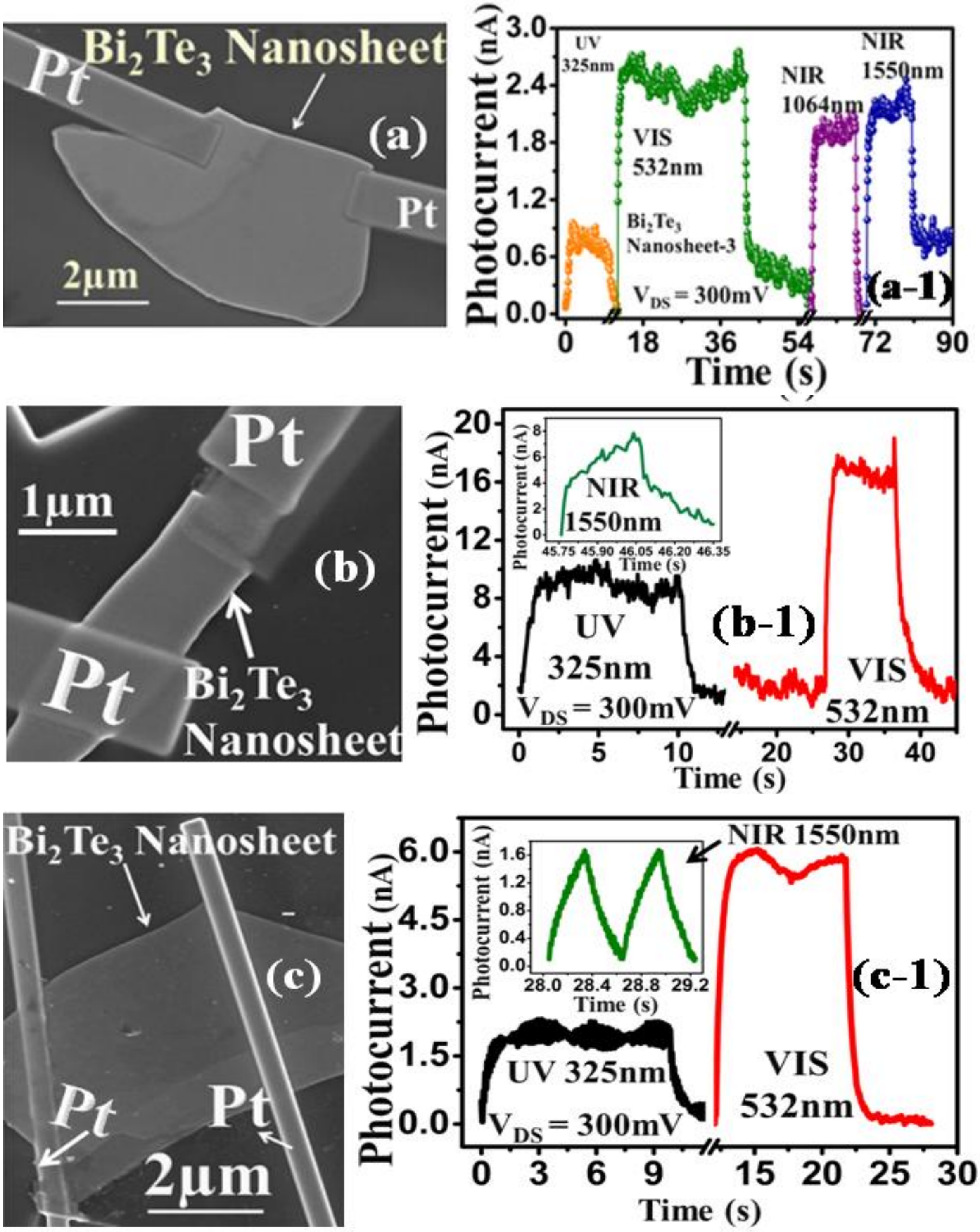

**Figure S2:** Photoconductivity measurements on three more nanosheet devices

### 3. Optoelectronic characterization of $Bi_2Te_3$ nanowire

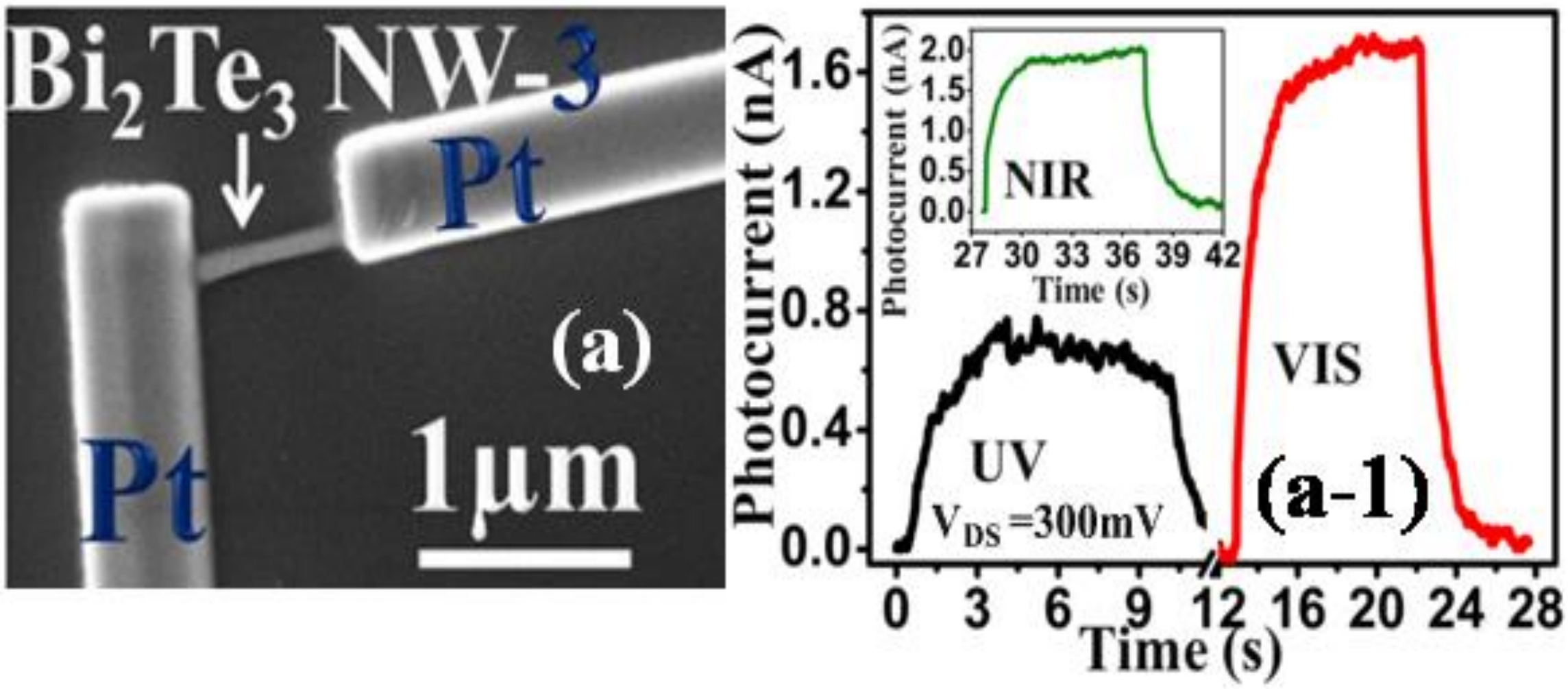

**Figure S3:** Photoconductivity measurements on one more fabricated nanowire device

## 4. Robustness and enhancements in photoconduction measurements

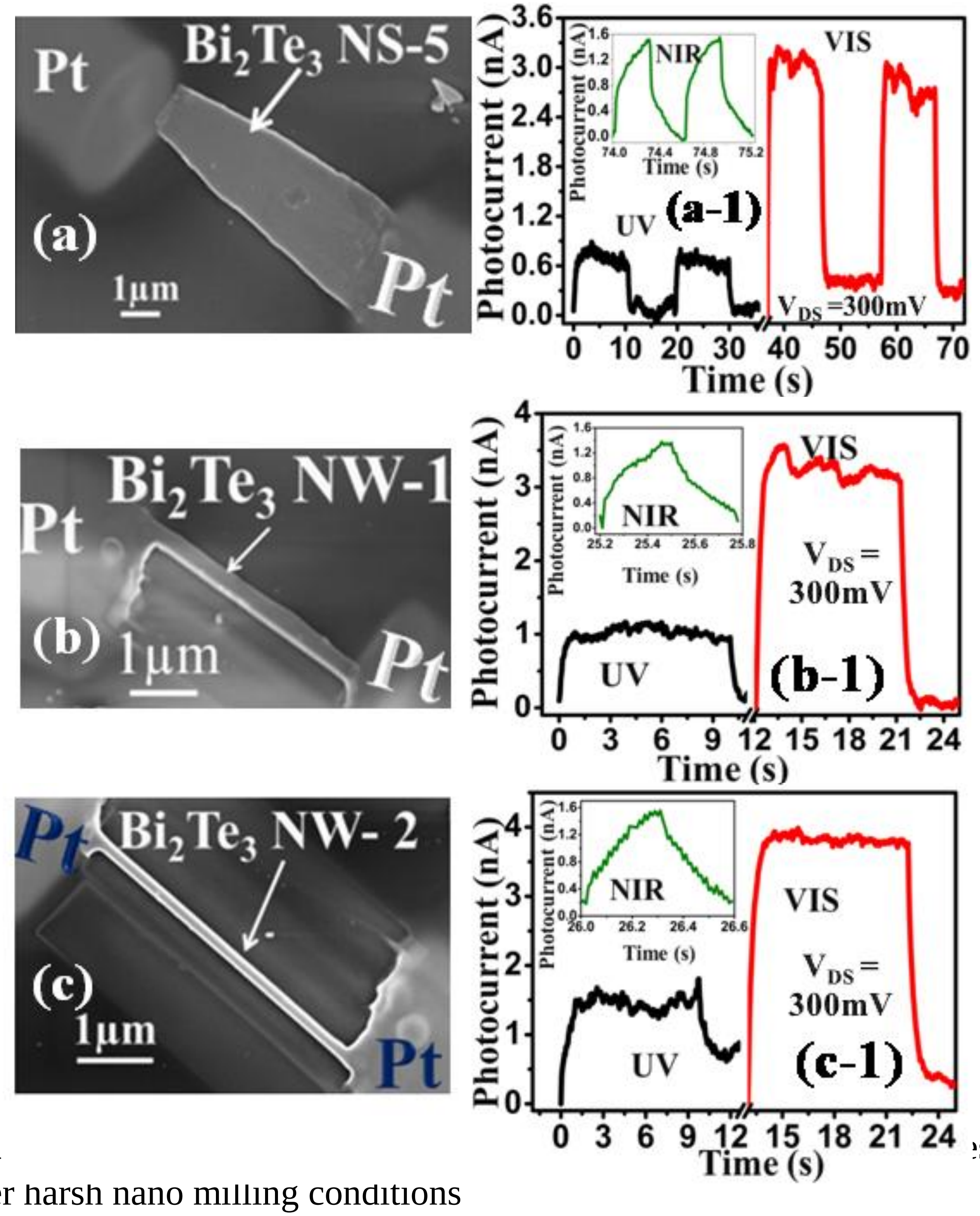

**Figu** ess
under harsh nano milling conditions

## 5. Photoconductivity of (visible) of NW1 device

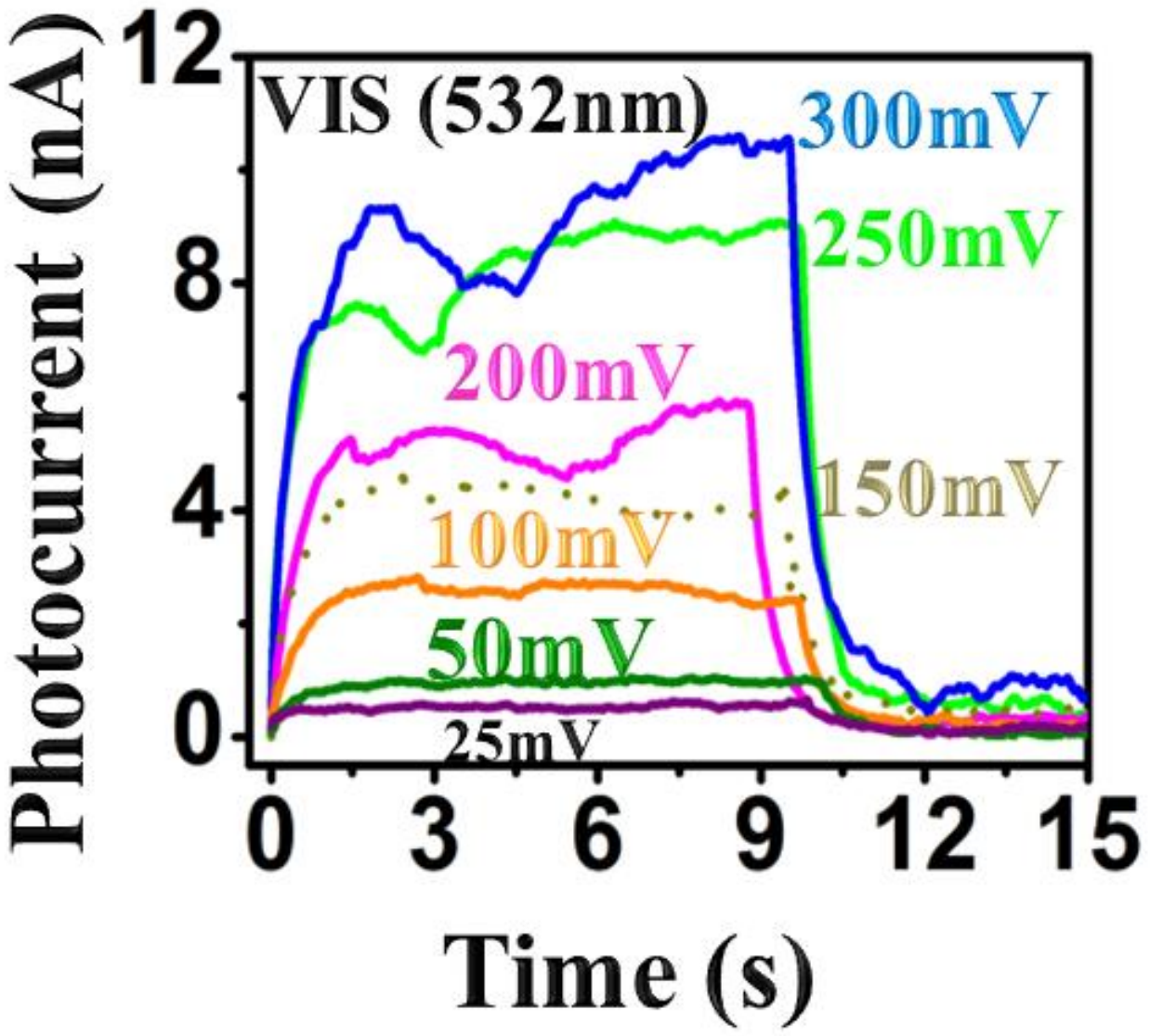

**Figure S5:** Bias voltage dependent photoconductivity measurements under illumination of visible light and device is NW1.

**6. Photoconductivity of (NIR) of NW2 device**

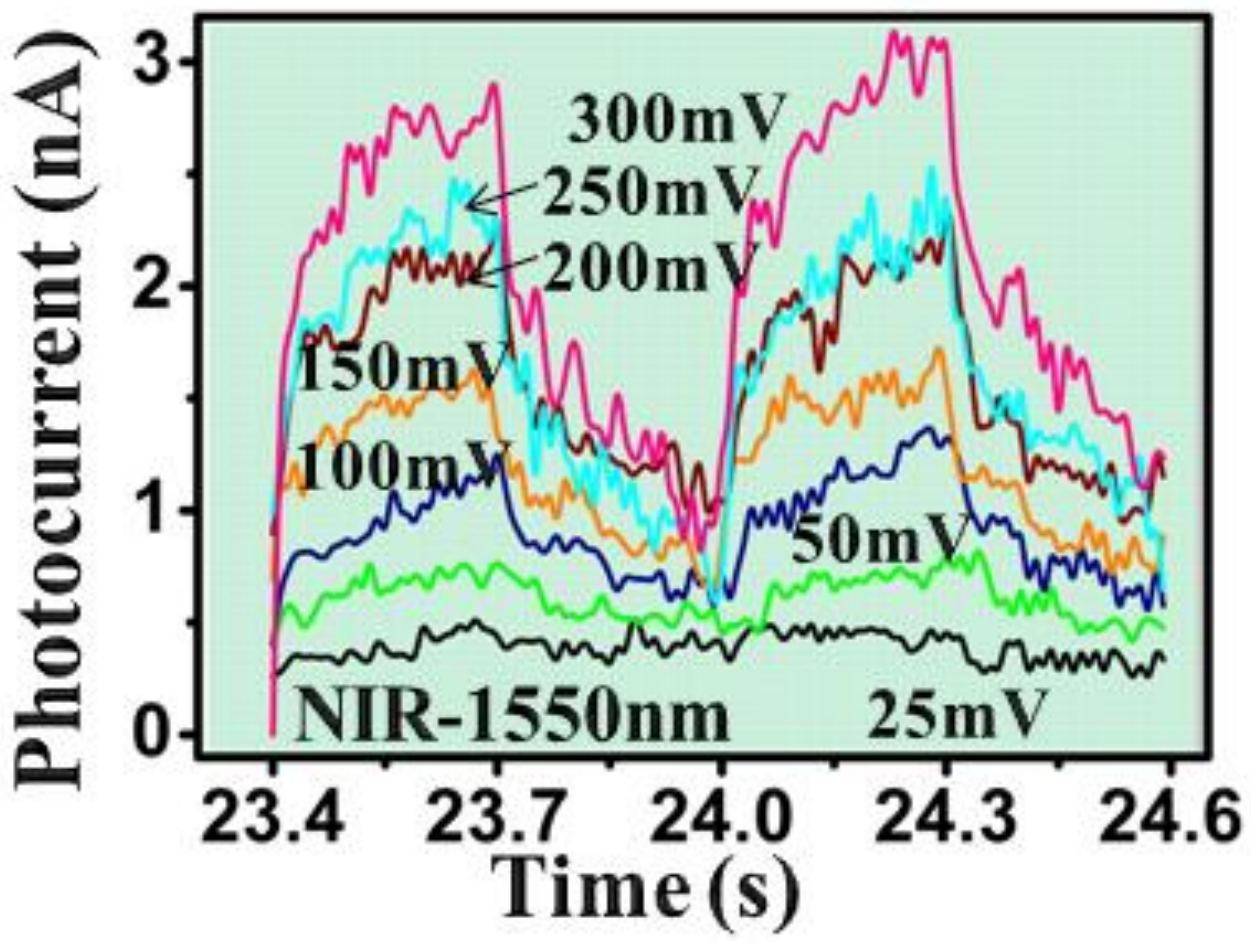

**Figure S6:** Bias voltage dependent photoconductivity measurements under illumination of NIR light and device is NW2.

## 7. Photoconductivity of (NIR) of Fab_NW1, Fab_NW2 and Fab_NW3 devices

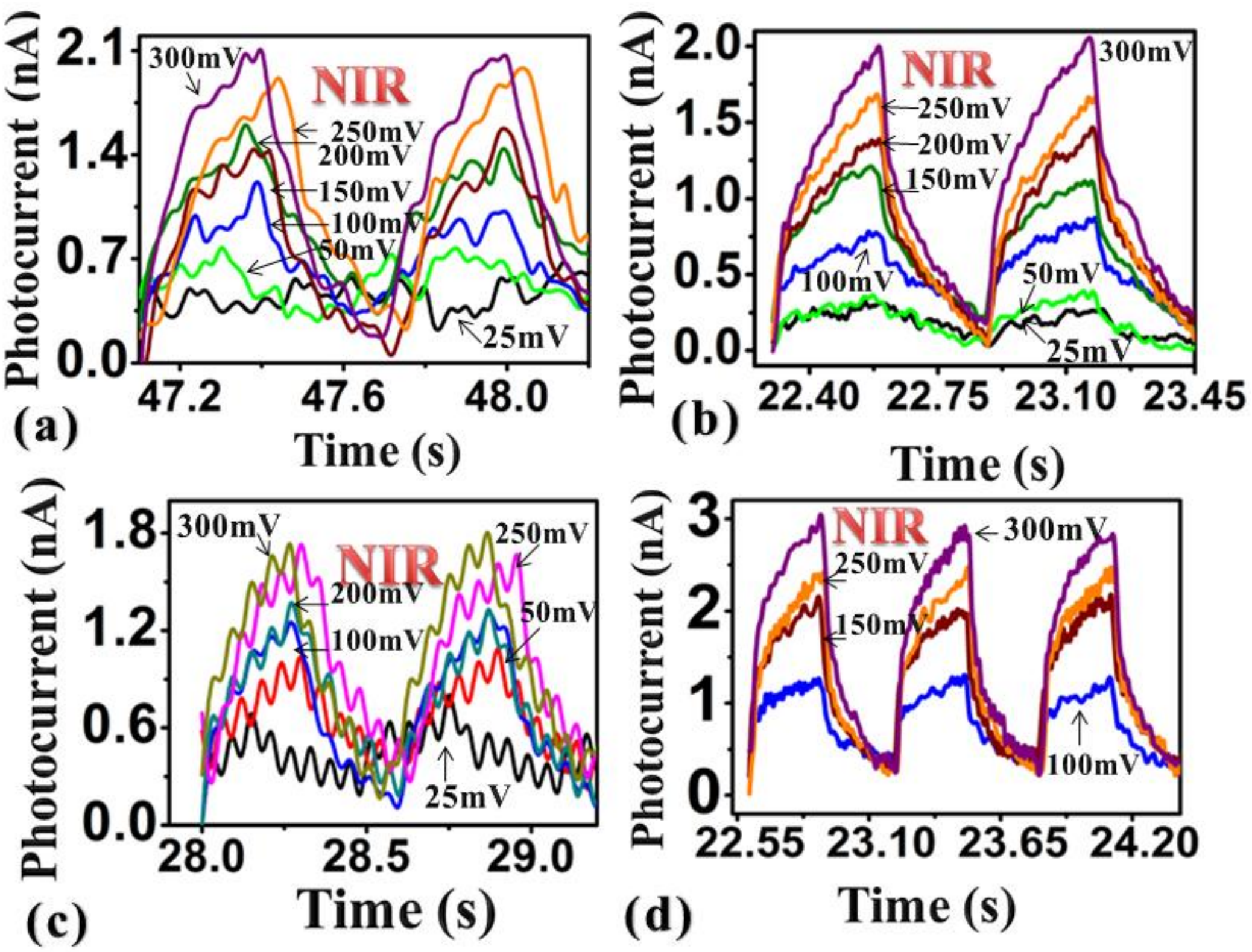

**Figure S7:** Bias voltage dependent photoconductivity measurements under illumination of NIR light and devices are Fab_ NW1, Fab_ NW2 and Fab_ NW3.

## 8. Photoconducting gain of the device (Fab_NW3)

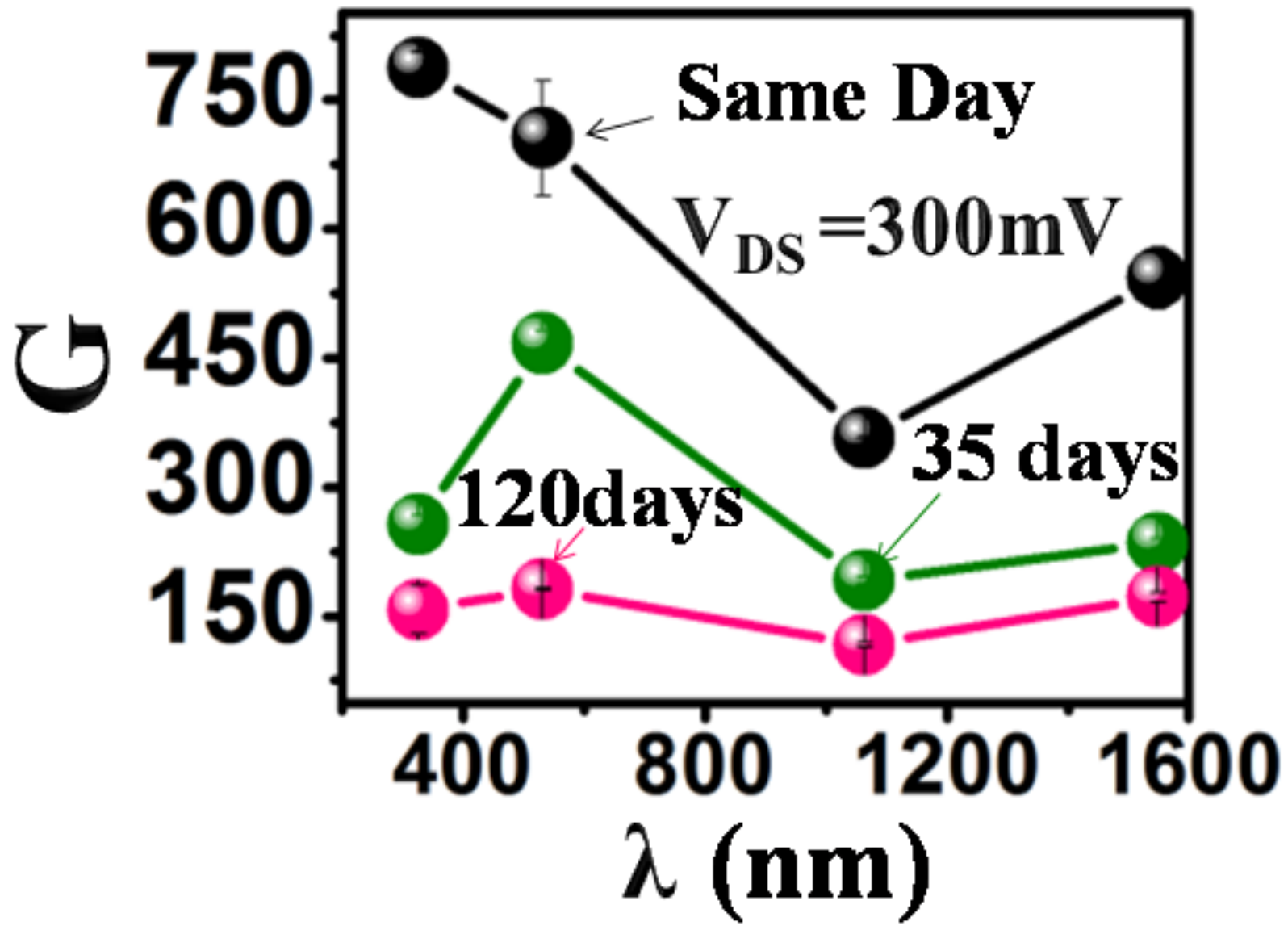

**Figure S8:** Photoconductive gain of the device Fab_NW3.